\documentstyle[preprint,prd,aps]{revtex}

\newcommand{\beq}{\begin{equation}}
\newcommand{\beqa}{\begin{eqnarray}}
\newcommand{\eeq}{\end{equation}}
\newcommand{\eeqa}{\end{eqnarray}}

\newcommand{\s}{{\sigma}}

\def\p{\partial}

\begin{document}

\draft
\preprint{gr-qc/9810002, UTAP-303}

\title{
Resolving the singularity of the
Hawking-Turok type instanton
}

\author{Takeshi Chiba\footnote{
JSPS research fellow.}
}

\address{
Department of Physics, University of Tokyo, 
Tokyo 113-0033, Japan
}
\date{\today}
\maketitle

\begin{abstract}
We point out that the singular instanton of Hawking-Turok
type, in which the singularity occurs due to the divergence of a
massless scalar field, 
can be generated by Euclideanized regular $p$-brane solutions in 
string (or M-) theory upon compactification to  four dimensions.
\end{abstract}


\section{introduction}

Hawking and Turok(HT) have recently discovered an instanton for the 
creation of an open universe from nothing\cite{ht}. This leads to an
interesting possibility of realization of open inflation with more
realistic form of the inflaton potential. However, the HT-instanton 
contains a genuine singularity: both the curvature and the scalar
field diverse there. Although the action is finite, further
justification for allowing such a singular instanton is necessary.
In fact, as Vilenkin has shown\cite{vilenkin}, instantons with the
same singularity structure as that of HT-instanton lead to
unacceptable physical consequences: the decay of flat space. 

Garriga has recently shown \cite{garriga} that the singularity can 
be resolved in five dimensions and that flat space with a compact extra
dimension can be sufficiently long lived if the size of the extra
dimension is large compared with the Planck length in four dimensions
(see also \cite{witten82}). 
In fact, the fundamental energy scale in M-theory\cite{mtheory} is the
eleven-dimensional Planck mass $M_{11}\simeq 10^{17}$GeV\cite{witten,bd}.
Unfortunately, Garriga's solution is not derived  in the context of
string(or M-)theory. 
On the other hand, Larsen and Wilczek have shown\cite{lw} that a class of
(3+1)-dimensional FRW cosmologies can be embedded within a variety of
solutions of string theory. 

In this note, being motivated by Larsen and
Wilczek's approach, we show that the instanton with the same 
singularity structure can be generated by Euclideanized regular
$p$-brane solutions in string theory (or eleven-dimensional 
supergravity) upon compactification to four dimensions.

\section{Higher-dimensional resolution of cosmological singularities}

\subsection{from five to four}

Let us briefly review the procedure by which Larsen-Wilczek embed a
FRW universe model in a  non-singular fashion. They started the
discussion by reference to the five-dimensional interior Schwarzschild
metric as an example. It can be written as
\beqa
ds^2&=&e^{-4\sigma}dy^2-e^{4\sigma}d\tau^2+\tau^2d\Omega_3^2,
\label{5d-metric}\\
&~&e^{-4\sigma}={\mu\over \tau^2}-1,\label{5d-sch}
\eeqa
where $\mu > 0$ is a mass parameter. The range of $\tau$ is
$0<\tau^2<\mu$. Compactification along $y$ induces an
effective four-dimensional dilaton $e^{-2\phi_4}=e^{-2\sigma}$ with
the string frame metric $g_{S\mu\nu}$ in which the $dy^2$ term is
omitted in Eq.(\ref{5d-metric}). In terms of the Einstein frame metric
$g_{E\mu\nu}=e^{-2\phi_4}g_{S\mu\nu}$, the four-dimensional cosmology
becomes
\beq
ds_E^2=-e^{2\sigma}d\tau^2+e^{-2\sigma}\tau^2d\Omega_3^2.
\label{4d-metric}
\eeq
We shall work in the Einstein frame line element because it is invariant
under the standard duality transformations in the string theory 
context and because a test particle follow geodesic in this frame\cite{lw}.
Since in the Einstein frame the effective four-dimensional action is
that of a massless scalar, minimally coupled to gravity, the
four-dimensional FRW cosmology is the closed model with matter of
$\rho=p$ and singularities at both the big-bang and the big-crunch.

Thus, the five dimensional vacuum black hole solution  with a
coordinate singularity (the horizon) and a physical singularity (the
origin) becomes four dimensional cosmology with a massless scalar field and
physical singularities at both initial and final times.

We note here that Garriga's five-dimensional regular instanton is
nothing but the five-dimensional Schwarzschild instanton. 
To show that, we introduce the time coordinate $t$ by 
\beq
dt=e^{2\sigma}d\tau,
\label{pro1}
\eeq 
so that $t^2=\mu-\tau^2$.\footnote{Here the integration constant is
neglected.} Then Eq.(\ref{5d-metric}) becomes
\beqa
ds^2&=&-dt^2+\tau^2d\Omega_3^2+e^{-4\sigma}dy^2,\\
&=&-dt^2+(\mu-t^2)d\Omega_3^2+{t^2\over \mu-t^2}dy^2.\label{5d-cosmic}
\eeqa
Thus, if we complexify $t$ and $y$ such that 
\beq
T=it,~~ Y=iy
\label{pro2}
\eeq
with $T$ and $Y$ being real, then Eq.(\ref{5d-cosmic}) is Euclideanized as
\beq
ds^2=dT^2+(\mu+T^2)d\Omega_3^2+{T^2\over \mu+T^2}dY^2,
\eeq
which is the same as the Garriga's regular five-dimensional instanton
solution\footnote{Garriga's instanton can also
be obtained from the exterior Schwarzschild metric.} 
(if $0\leq Y \leq 2\pi \mu^{1/2}$) and solves the Euclidean 
field equations for pure gravity in 
five dimensions\cite{garriga}.
 Changing into the Einstein frame after 
compactifying $Y$ yields the following metric and massless scalar field:
\beqa
ds^2&=&{T\over \sqrt{\mu+T^2}}\left(dT^2+(\mu+T^2)d\Omega_3^2\right)\\
&\equiv& d\bar{T}^2+b(\bar{T})^2d\Omega_3^2,\\
\phi& \propto & -\ln \left({T^2\over \mu +T^2}\right).
\eeqa
The internal radius of the fifth dimension can be zero, but it is just 
a coordinate singularity. However, that zero-point induces a
singularity of the Einstein frame metric in four dimensions. In that
way, the singularity in the four-dimensional world is induced. 
Near the singularity $T\simeq 0$, $\bar{T}-\bar{T}_f \simeq
2\mu^{-1/4}T^{3/2}/3$ with $\bar{T}_f$ being a constant. Thus the
scale factor $b(\bar{T})$ and the scalar field behave as
\beqa
b(\bar{T}) &\simeq & \mu^{1/4}T^{1/2} \propto |\bar{T}-\bar{T}_f|^{1/3},\\
\phi &\propto & -\ln T +{\rm const.} \propto -\ln |\bar{T}-\bar{T}_f|
+{\rm const.}
\eeqa
This is the same singularity structure as that of the Hawking-Turok(or
Vilenkin) instanton\cite{ht,vilenkin,garriga}.

\subsection{from ten to four}

Larsen-Wilczek further generalized the model and embedded the cosmology in a
solution that originates from string theory. 

The starting point is the classical solution associated with the
black $p$-brane\cite{gs,dlp,ct}. In the string frame it is written as
\beqa
ds_S^2&=&e^{2\xi}(e^{-4\s}dy^2+dx_1^2+\dots +
dx_p^2)+e^{-2\xi}(dx_{p+1}^2+\dots+dx_5^2)\nonumber\\
&~&+e^{-2\xi}(-e^{4\s}d\tau^2+\tau^2d\Omega^2_3),\\
e^{-2\phi_{10}}&=& e^{-2\xi(p-3)},\\
F_{p+2}&=&\p_{\mu}e^{4\xi}dy\wedge dx_1\wedge y \dots \wedge dx_p
\wedge dx^{\mu},
\eeqa
where $F_{p+2}$ is the $(p+2)$-form Ramond-Ramond field strength and 
$\xi$ and $\s$ are defined as
\beqa
e^{-4\xi}&=&{q\over \tau^2}+1,\\
e^{-4\sigma}&=&{\mu\over \tau^2}-1.
\eeqa 
Compactification to four-dimensions\footnote{Here we only consider
toroidal compactification. For sphere compactification, see
\cite{bdlps}.} induces an effective dilaton
$e^{-2\phi_4}=e^{2\xi-2\s}$ independent of $p$. Remarkably, the four 
dimensional
Einstein metric is again Eq.(\ref{4d-metric}). Thus the black p-brane 
generates the same four-dimensional cosmology as the five-dimensional 
Schwarzschild solution does. 

The corresponding Euclidean solution can be immediately obtained 
by following the same procedure ({\ref{pro1}) and (\ref{pro2}) made
in five-dimensional case. Namely, we have
\beqa
ds_S^2&=&\left({\mu+T^2\over q+\mu +T^2}\right)^{1/2}\left({T^2\over \mu
+T^2}dY^2+dx_1^2+\dots + dx_p^2\right)+
\left({\mu+T^2\over q+\mu +T^2}\right)^{-1/2}(dx_{p+1}^2+\dots+dx_5^2)
\nonumber\\
&~&+\left({\mu+T^2\over q+\mu +T^2}\right)^{-1/2}
(dT^2+(\mu+T^2)d\Omega^2_3).
\eeqa
This solution is again regular if $0\leq Y \leq 2\pi
\left((q+\mu)\mu\right)^{1/4}$.

\subsection{from ten to eleven}

One solution in Type IIA  string theory is derived from another in 
eleven-dimensional supergravity by compactifying on a
circle\cite{dfis,guven,kko}: 
\beqa
\exp(4\phi_{10}/3)&=&g_{M 11,11},\\
g_{S\mu\nu}&=&\exp(2\phi_{10}/3)g_{M \mu\nu},
\eeqa
where $g_{M \mu\nu}$ is M-theory frame metric.
Conversely, one solution in Type IIA  string theory is 
``oxidated'' to another solution in eleven-dimensional supergravity. For
example, black five-brane solution in eleven-dimensional supergravity
can be obtained from black four-brane in string theory:
\beqa
ds_{11}^2&=&e^{4\xi/3}(e^{-4\s}dy^2+dx_1^2+\dots +
dx_5^2)+e^{-2\xi}dx_6^2\nonumber\\
&&+e^{-2\xi}(-e^{4\s}d\tau^2+\tau^2d\Omega^2_3).
\eeqa
We also get black two-brane in  eleven dimensions from black
two-brane in ten dimensions:
\beqa
ds_{11}^2&=&e^{8\xi/3}(e^{-4\s}dy^2+dx_1^2+dx_2^2)+
e^{-4\xi/3}(dx_3^2 + \dots + dx_6^2)\nonumber\\
&&+e^{-4\xi/3}(-e^{4\s}d\tau^2+\tau^2d\Omega^2_3).
\eeqa
Corresponding Euclidean solutions can be obtained in the same manner. 

\section{summary}

We have shown that the singularity of the instanton of Hawking-Turok
type can be resolved from higher-dimensional viewpoint within the
context of M-(or string) theory. 
We have also shown that Garriga's instanton is nothing but 
the five-dimensional Schwarzschild instanton. 
The same singularity 
structure emerges when a regular instanton in higher dimensions is 
compactified to four dimensions. 
The fundamental length scale in M-theory is much larger than the four
dimensional Planck length, and therefore the decay of flat space is
strongly suppressed.

\acknowledgments
A part of this work was reported at the Takehara meeting on general
relativity (July, 27-30, 1998) held at the old building of the Institute 
formerly known as Research Institute for Theoretical Physics in 
Hiroshima. 
The author would like to thank Takahiro Tanaka and Misao Sasaki for
useful discussions and comments. 
This work was supported by JSPS Research Fellowships for 
Young Scientists.

\end{document}